\documentclass[preprint]{aastex6}
\usepackage{graphicx}
\usepackage{epsfig}
\usepackage{multirow}
\usepackage{footnote}
\usepackage{url}

\shorttitle{Dust and Companions around V488 Per}
\shortauthors{S. Sankar {\it et al.}}

\begin{document}

\title{\large \bf V488 Per revisited: no strong mid-infrared emission features and no evidence for stellar/sub-stellar companions}

\author{\large Swetha Sankar}
\affil{email: swethasnkr@ucla.edu \\
Department of Physics and Astronomy, University of California, Los Angeles, CA 90095-1562, USA}

\author{\large Carl Melis}
\affil{email: cmelis@ucsd.edu \\
Center for Astrophysics and Space Sciences, University of California, San Diego, CA 92093-0424, USA} 

\author{\large Beth L.\ Klein}
\affil{Department of Physics and Astronomy, University of California, Los Angeles, CA 90095-1562, USA}

\author{\large B.\ J.\ Fulton}
\affil{NASA Exoplanet Science Institute / Caltech-IPAC, Pasadena, CA 91125, USA}

\author{\large B.\ Zuckerman}
\affil{Department of Physics and Astronomy, University of California, Los Angeles, CA 90095-1562, USA}

\author{\large Inseok Song}
\affil{Department of Physics and Astronomy, University of Georgia, Athens, GA 30602-2451, USA}

\and

\author{\large Andrew W.\ Howard}
\affil{California Institute of Technology, Pasadena, CA 91125, USA}

\begin{abstract}

\large{
We present characterization of the planetary system architecture for
V488 Per, the dustiest main sequence star known with a fractional infrared luminosity
of $\approx$16\%.
Far-infrared imaging photometry confirms the existence of an outer planetary system
dust population with blackbody-fit temperature of $\approx$130\,K.
Mid-infrared spectroscopy probing the previously-identified $\approx$800\,K 
inner planetary system dust population does not detect any obvious solid-state emission features,
suggesting either large grain sizes that mute such emission and/or grain compositions
dominated by species like amorphous carbon and metallic iron which do not produce such features.
In the latter case, the presence of significant quantities of iron-rich material could be indicative of
the active formation of a Mercury-like planet around V488 Per.  In any event, the absence of solid-state 
emission features is very unusual among main sequence stars with copious amounts of warm orbiting 
dust particles; we know of no other such star whose mid-infrared spectrum lacks such features.
Combined radial velocity monitoring and adaptive optics imaging
find no evidence for stellar/sub-stellar companions within several hundred AU
of V488 Per.
\\*[1.0mm]}

\end{abstract}

\begin{figure}
 \centering
 \includegraphics[width=160mm]{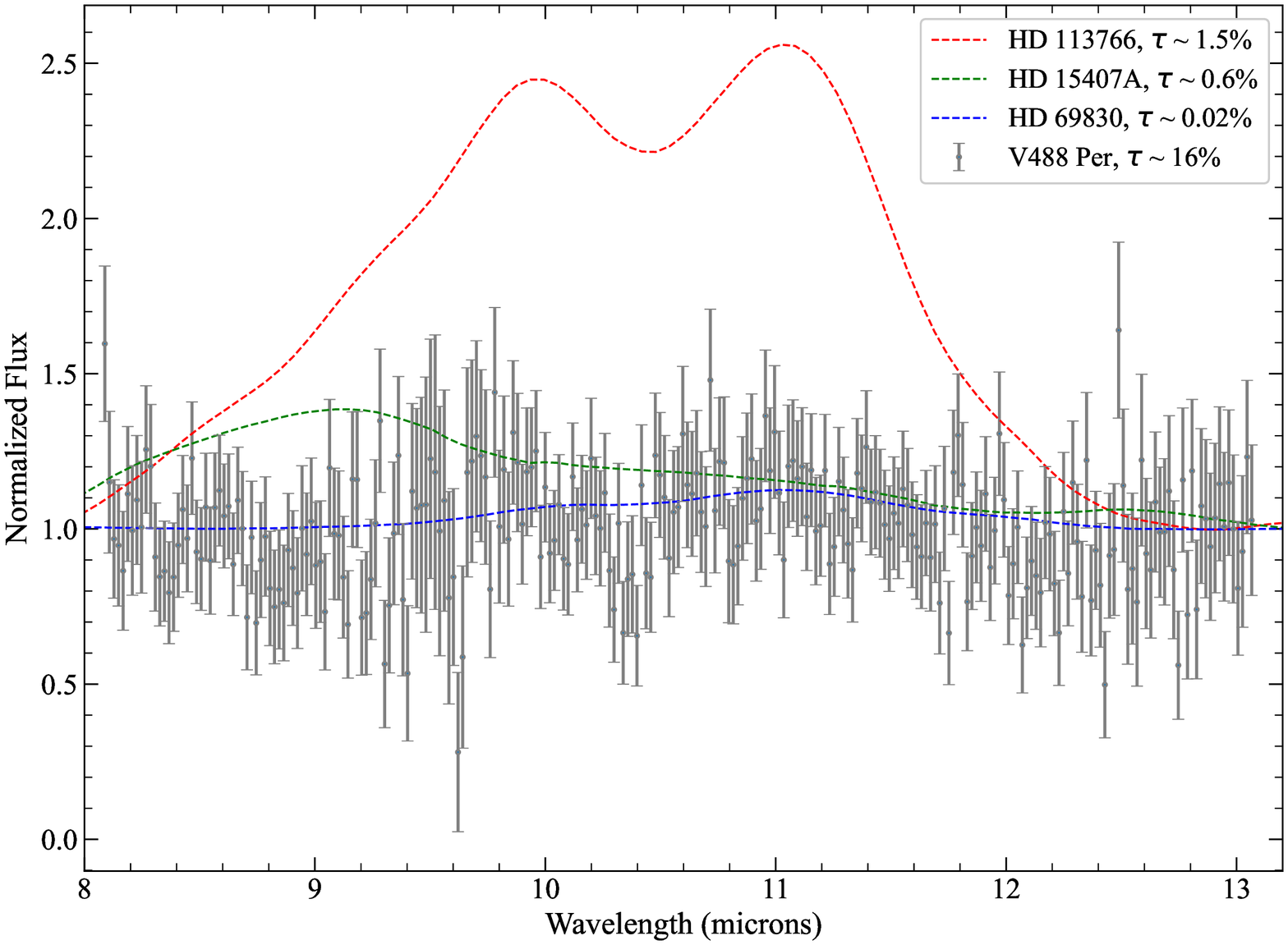}
 \caption{\label{fig:compare} \large{Mid-infrared spectra of V488 Per and 
              exemplar main sequence stars with
              warm inner planetary system dust. Each spectrum
              is normalized by a blackbody model fit to regions of the spectrum that
              do not have solid-state emission features, typically near 7.8\,$\mu$m and 13.5\,$\mu$m.
              The signal-to-noise ratio per pixel for the exemplar spectra range between roughly 100-200
              and as such the error bars for those spectra are comparable to the thickness of the
              lines used to plot them.
              The fractional infrared luminosity ($\tau$$=$L$_{\rm IR}$/L$_{\rm bol}$) 
              of each source shown is given in the legend.
              HD\,113766 (e.g., \citealt{chen06}; \citealt{lisse08})
              is an exemplar for amorphous and crystalline solid-state emission features seen in
              stars with fractional infrared luminosities $\gtrsim$1\%; such sources present a range
              of solid-state emission feature peak strengths of 2-5$\times$ the continuum level.        
              HD\,15407A (e.g., \citealt{Melis_2010}; \citealt{fujiwara12dust}) is an exemplar
              for silica-dominated disks around main sequence stars; such sources present
              solid-state emission feature peak strengths $\approx$1.5$\times$ the continuum level.
              HD\,69830 (e.g., \citealt{beichman05}; \citealt{lisse07})
              is an exemplar for the types of emission features seen around main sequence stars
              with lower fractional infrared luminosities ($\sim$0.01\%). 
              Section \ref{sec:dust} provides more discussion.
              }}
\end{figure}

\large{

\section{\large \bf Introduction}
\label{sec:intro}

Many main sequence stars are now known to host substantial quantities of dusty 
material in their inner planetary systems 
(e.g., \citealt{melis16}; \citealt{moor21}; \citealt{melis21}; \citealt{absil21}, and references therein). 
Where this inner planetary system dust comes from, what drives its evolution, 
and what its fate ultimately will be are matters of active research. 
Since this dust could be indicative of processes that might influence the
formation, evolution, and habitability of Earth-like planets, it is essential to develop a 
firm understanding of it. 
Systematic investigation of the dust in these systems and 
their greater planetary system architecture will provide insight into physical processes at 
play: e.g., how dust is generated and removed 
and the impact of dynamical perturbers on the origin and evolution of the dust.

Exceptionally dusty main sequence stars are frequently found to reside in 
multiple stellar systems raising the question of whether or not multiplicity and 
chaotic dynamics contribute to generating these 
systems (e.g., \citealt{zuckerman15}; \citealt{moor21}). 
For two older ($>$1 Gyr) systems with mid-infrared excess emission
indicative of inner planetary system dust,
HD\,69830 and BD+20$^{\circ}$~307, 
it does indeed seem like additional planetary or stellar
companions play a role in either destabilizing 
or corralling dust-producing rocky bodies leading to the unusually high levels of
dust (e.g., \citealt{zuckerman08}; \citealt{payne09}).

\citet{Nesvold_2016} study how a stellar-mass perturber orbiting exterior to 
and inclined to a planetesimal disk can excite the disk by the
Kozai-Lidov mechanism resulting in dust-producing collisions.
\citet{moor21} discuss how a widely orbiting stellar companion may act via
Kozai-Lidov mechanisms to send
outer planetary system bodies on inner planetary system-crossing orbits
where they may release dust via disruption through sublimation or collisions.

Planetary mass companions orbiting a star may undergo a dynamical instability
similar to the Late Heavy Bombardment hypothesized to have occurred in
the solar system (e.g., \citealt{gomes05}). \citet{bonsor13,bonsor14} investigate
such scenarios numerically, finding that sustained dust production in the inner
planetary system can only be achieved when a planet migrates into a planetesimal
belt.
\citet{fujiwara09,fujiwara12} explore such a possibility in two stars with
dusty inner planetary systems, rejecting it as an explanation for HD\,15407A
due to the lack of a detectable outer planetary system disk.
However, \citet{lisse12} strongly suggest such a model is appropriate for $\eta$ Crv
based on a detected outer planetary system disk and possible water-ice features
seen in its disk mid-infrared spectrum.
\citet{melis21} invoke instabilities in tightly-packed inner planetary systems of
small planets (Earth to Neptune-sizes) to produce the observed very dusty
main sequence stars.

Collisionally-produced dust particles with sizes $\lesssim$1\,$\mu$m
typically generate solid-state emission features around 10\,$\mu$m.
Characterization of solid-state emission features
can provide insight into where the dust in extreme debris disks originates from
and what physical processes play a role in its evolution (e.g., 
\citealt{honda04};
\citealt{song05};
\citealt{chen06};
\citealt{lisse07,lisse08,lisse09,lisse12,lisse17,lisse20};
\citealt{rhee07,rhee08};
\citealt{currie11};
\citealt{johnson12};
\citealt{olofsson12};
\citealt{melis13};
\citealt{ballering14};
\citealt{morlok14};
\citealt{mittal15}).
For example, highly energetic impacts that result in the production of crystalline silicates or 
silica (e.g., \citealt{lisse09}) would be suggestive of dynamically hot populations
of mature rocky bodies pointing to the need for eccentricity excitation mechanisms.

Thus far, most extreme debris disk stars have mid-infrared spectral features 
from silicate species indicating
large quantities of highly processed material (crystalline and silica grains) in addition to
less processed amorphous species (e.g., Figure \ref{fig:compare}).
However, it is worth noting three unusual systems that might be host to rather
different dusty material (\citealt{melis13}; \citealt{lisse17}). 
Systems with atypical dust properties could provide
information about unique dust stoichiometry and hence composition or dust production processes.
Thorough characterization and modeling of mid-infrared spectral features 
for the dustiest main sequence stars is essential in developing a
complete picture for the types of compositions at a mineralogical level these systems
host and the dynamical conditions that lead to their production through collisional processes
(e.g., \citealt{lisse09,lisse12}; \citealt{meng14}; \citealt{su20}).

We seek to obtain constraints on the overall planetary system architecture for the dustiest
main sequence stars known.
V488 Per, a Solar-type star belonging to the $\approx$80\,Myr old $\alpha$ Persei cluster
\citep{soderblom14}, is of particular interest 
as it hosts what is currently the largest known fractional infrared luminosity for a main sequence
star of
$\tau$($=$L$_{\rm IR}$/L$_{\rm bol}$)$\approx$16$\%$ \citep{Zuckerman_2012}.
Basic stellar properties for V488 Per are given in Table \ref{tab:V488Per_param}.
\citet{Zuckerman_2012} suggest that V488 Per is possibly host to two separate belts of
dust with temperatures of $\approx$820\,K and 120\,K, but the basis for the cooler
dust component was a single {\it WISE} excess measurement at 22\,$\mu$m.

In this paper, we present new data that further characterize the dust and planetary
system architecture of V488 Per. Section \ref{sec:obs} details the new data sets,
Section \ref{sec:disc} discusses analyses based on measurements from the data,
and Section \ref{sec:conclusions} discusses possible interpretations and future observations
to investigate them.
 
\section{\large \bf Observations}
\label{sec:obs}

\subsection{\large \bf Optical Spectroscopy}

Optical echelle spectra for V488 Per and a radial velocity standard of similar temperature class, 
HR 124 \citep{nidever02}, were collected at Lick Observatory 
with the Automated Planet Finder (APF) telescope and Levy spectrograph \citep{vogt14}, as well as the Shane telescope and Hamilton spectrograph (\citealt{vogt87}; \citealt{pakhomov13}).
We obtained these data to search for doppler signatures from
any companions to V488 Per at separations that could not be probed with imaging techniques;
the iodine cell was not used due to the optical faintness of V488 Per for these telescopes (Table \ref{tab:V488Per_param}).
Two-dimensional raw frames are bias subtracted, flat-fielded, extracted into one-dimensional
spectra through straight summing of flux inside an aperture defined for each individual echelle order, 
and finally wavelength calibrated with comparison lamp spectra.
Radial velocities are measured through cross-correlation with HR 124 and corrected
to the heliocentric reference frame.
Table \ref{tab:obs} reports observation epochs, data quality, and measured velocities
while Figure \ref{fig:MassLimit} presents measurements and associated analysis.

\subsection{\large \bf {\it TESS}}

To help assess the stellar inclination angle and hence better interpret doppler
mass limits by removing the sin$i$ ambiguity, we downloaded and analyzed
{\it TESS} lightcurve data for V488 Per and other $\alpha$ Per stars with similar
spectral type and spectroscopic projected rotational velocities \citep{Stauffer_1985}.
All lightcurves are
generated from Full Frame Imaging data acquired in Sector 18 from 2019 Nov 03 to 2019 Dec 16.
Data products are MIT Quick Look Pipeline (QLP; \citealt{Huang_2020_a, Huang_2020_b})
lightcurves as downloaded from MAST. Some times with bad data are removed by hand
before analysis. Figure \ref{fig:TessLC} shows {\it TESS} lightcurves used for this work.

\subsection{\large \bf Adaptive Optics}

Diffraction-limited thermal-infrared adaptive optics (AO) imaging of V488 Per 
was obtained at Keck Observatory with the facility AO system \citep{wizinowich00,wizinowich13}
and NIRC2 camera on UT 2013 Feb 03. We conducted observations using
the L$'$ filter (central wavelength and bandpass 
of 3.776$\pm$0.700\,$\mu$m) and the narrow camera
setting, resulting in a plate scale of 0.009952\,$'$$'$\,pixel$^{-1}$ \citep{yelda10}.
We obtained these data to further rule out possible contamination of V488 Per
by a background object and then to perform a shallow search for possible
companion stars. The excess emission level for V488 Per at a wavelength
of $\approx$3.8\,$\mu$m is just over twice the stellar photosphere emission level
(Figure \ref{fig:SED}), thus
any contaminating object that would be responsible for the apparent infrared excess emission
toward V488 Per would need to be of comparable brightness as V488 Per (the star) alone.

Seven dithered exposures of 3.0~seconds each 
(0.3~seconds integration $\times$ 10 coadds) were obtained resulting in a total integration
time of 21~seconds. Background-subtracted and flat-fielded individual images are registered, then
combined to produce a final image for analysis which is displayed in Figure \ref{fig:MassLimit}.

\subsection{\large \bf COMICS Spectra}

We collected COMICS (\citealt{kataza00}; \citealt{okamoto03}) R$\approx$170
mid-infrared spectroscopic data at the Subaru 8.2\,m Telescope on UT 06 Nov 2014.
The night featured seeing in the mid-infrared of $\approx$0.4$'$$'$
and a precipitable water vapor value of $\sim$4\,mm.
Observations and data reduction follow \citet{su20},
important aspects of observations for V488 Per are discussed here.

The observing strategy for V488 Per started with observations of 
the calibrator star HD\,21552 (K3\,III), a sequence on V488 Per 
lasting $\approx$90~minutes of wall-clock time, 
a second visit to HD\,21552 followed by another $\approx$90~minute sequence
of observations on V488 Per, then a final visit to HD\,21552.
All exposures for both stars were chopped in an ABAB pattern keeping all beams on slit.
V488 Per is faint for ground-based mid-infrared astronomy (Figure \ref{fig:SED}), 
so each saveset obtained for
it had 222 total exposures or 111 chop pairs, the accumulation of which was sufficient to see
the spectral signal. By comparison, the bright calibrator only needed 24 total exposures
or 12 chop pairs to obtain high ($\sim$50) signal-to-noise.
During reduction, each chop pair is differenced
and examined for quality. Those chop pairs showing strong residual
structure across the detector (indicating rapidly varying background conditions)
are discarded.
Chop pairs that are kept are then rectified along the spatial axis such that night sky 
emission lines run vertically only (initially they are slightly tilted horizontally).
A second pass of background removal is then done by subtracting at each dispersion pixel location
the median value of background-only pixels along the spatial axis.
At the end of assessment, roughly two-thirds of the first spectral sequence on V488 Per
was kept and roughly one-third of the second spectral sequence.

Fully processed chop pairs in each sequence
are median-combined and the positive and negative spectral beams are extracted
via a straight aperture sum.
Uncertainties on spectral samples are calculated by determining the rms of background-only
pixels and summing that in quadrature for the number of pixels in the aperture and the Poisson
noise on the total summed flux in the aperture.
This is done for each of the two target spectral sequences and for spectra from each visit 
to the calibration star.

For each visit to the calibrator we
scaled and combined positive and negative beam spectra with a weighted mean, 
then used one spectrum from the set of three visits to correct the telluric
absorption in each positive and negative spectrum for
V488 Per. Telluric correction is done by dividing a 
science spectrum
by one of the calibrator visit spectra shifted in wavelength to provide the best cancellation of the strong
$\sim$9.5\,$\mu$m ozone featue. For each science spectrum, 
we selected the calibrator spectrum that resulted in the lowest rms residuals from telluric features;
in practice this was the calibrator spectrum obtained in the middle of the two sequences
on V488 Per.

Corrected science target positive and negative beam spectra in each sequence were then
scaled and combined via weighted mean. A wavelength scale is determined
based on a low-order polynomial fit to the position of known
bright sky emission lines. A Rayleigh-Jeans slope is used as an approximation
for the calibrator spectral shape and applied to the corrected science target spectra
to arrive at relative flux-calibrated spectra (i.e., the spectral
shape is robust, but the absolute flux level is not).

To obtain an accurate absolute flux scaling for the spectra for each sequence 
we measured photometry from images of the target and calibrator.
A short imaging acquisition sequence with the N12.4 filter (bandpass of 12.4$\pm$1.2\,$\mu$m) 
was obtained for V488 Per and the calibrator star HD\,21552
in chopping-only mode. Chop pairs were differenced to remove background emission
and aperture photometry performed on the science target and calibrator star.
We note that only a single point source was seen in the V488 Per field.
With an adopted flux density for HD\,21552 of 9.11\,Jy (extrapolated from the {\it WISE}-measured
11.56\,$\mu$m flux density of 10.48$\pm$0.11\,Jy), we measure
a 12.4\,$\mu$m flux density for V488 Per of 60$\pm$16\,mJy. Since this is consistent within the
errors with the {\it WISE} 11.56\,$\mu$m flux measurement for V488 Per of 41.0$\pm$0.6\,mJy, 
we adopt the {\it WISE} value to set the spectrum absolute flux level.
These absolute flux scales are accurate
at the $\approx$10\% level (this error source is not included in the spectral uncertainties).

A final N-band spectrum for V488 Per is then obtained by averaging the two separate sequences
together. The final spectrum signal-to-noise level is $\approx$6 near 11.5\,$\mu$m.

\subsection{\large \bf {\it Herschel} Imaging}

Simultaneous far-infrared observations at at 70 and 160\,$\mu$m were obtained with the 
ESA {\it Herschel} Space Observatory \citep{pilbratt10}
using the PACS imaging photometer \citep{poglitsch10}.
V488 Per was observed on UT 2012 Sep 11 (Proposal OT2\_cmelis\_3;
OBSIDs 1342250847 and 1342250848) in Mini Scan map mode.
Scan legs of 3$'$ length were observed at medium speed with two orientation angles
of 70 and 110$^{\circ}$ in the array coordinates.
High level image products are obtained from the {\it Herschel} Science Center
and aperture photometry is performed on these images as described
in \citet{Vican_2016}. We obtain a 70\,$\mu$m flux density in agreement
with the {\it Herschel} PACS Point Source catalog \citep{herschel20} and adopt their value;
we report for V488 Per a 70\,$\mu$m flux density of 66.5$\pm$6.5\,mJy and 160\,$\mu$m
flux density of 22$\pm$10\,mJy.

\section{\large \bf Analysis and Discussion}
 \label{sec:disc}
 
\subsection{\large \bf Search for Companions}

Figure \ref{fig:MassLimit} reveals the overall stability of the radial velocity measurements
made at Lick Observatory. A periodogram search does not reveal any signals hidden
in the noise, and with no obvious signal present in the data we proceed to estimate
mass sensitivity limits.
Radial velocity measurements given in Table \ref{table:radvelobs} are also
consistent with values 
previously measured for V488 Per \citep{Stauffer_1985,Mermilliod_2008,Zuckerman_2012}.
The measurement from \citet{Mermilliod_2008}, taken on HJD 2449655.560 (UT 1994 Oct 30) 
and having a value of $-$0.31$\pm$0.15\,km\,s$^{-1}$, is incorporated into our companion sensitivity
analysis. The other two measurements are not utilized as they
either did not have information regarding time of observation \citep{Stauffer_1985} 
or because of inaccuracies resulting from a much later spectral type radial velocity standard 
used for cross correlation \citep{Zuckerman_2012}.

To estimate mass sensitivity limits as a function of orbital separation
we utilized {\sf The Joker} \citep{Price-Whelan_2017} and adopted a stellar mass of 
0.84\,M$_{\odot}$ (Table \ref{tab:V488Per_param}).
{\sf The Joker} is designed to characterize a two body system with the requirement 
that the source behaves similar to a single-lined spectroscopic binary and exhibits 
variability in radial velocity measurements. It takes the input of multi-epoch radial velocity 
data with errors and by Monte-Carlo analysis produces a suite of possible fits for the data.
We conducted a set of {\sf The Joker} runs with a uniform prior distribution covering
a variety of orbital period ranges; 
in all runs, priors on the velocity semi-amplitude and the systemic velocity
were taken to be Gaussians with $\sigma$ of 3\,km\,s$^{-1}$ 
and 75\,km\,s$^{-1}$, respectively.
Discrete orbital period range bins are constructed out to a maximum period of
60,000~days ($\approx$30\,AU for a circular orbit).
For each run, 10$^3$ priors are generated but only a subset survive
the rejection sampling step of {\sf The Joker}.
Figure \ref{fig:MassLimit} presents some example orbital fits that are consistent with
the available radial velocity data.
Five runs per period range bin are executed and a mass sensitivity for that bin is calculated
from the median of the maximum masses obtained from each of the five runs.
The resulting mass-period sensitivity curve from these samplings
is displayed in Figure \ref{fig:MassLimit}; it is noted that mass sensitivities obtained
in this way are $m$sin$i$.

Radial velocity mass limits become increasingly insensitive with orbital period, especially
beyond the maximum time baseline covered as shown in Figure \ref{fig:MassLimit}.
To supplement mass sensitivities at larger orbital separations we make use
of the AO data. 
One other star was observed with the L$'$ filter on the same night as V488 Per,
but at a different time and significantly different airmass and as such
point-spread-function (PSF) subtraction was not fruitful. However, in
general the PSF of V488 Per matched that of the other observed star indicating a single
source.
Figure \ref{fig:MassLimit} shows what we conclude is a single, diffraction-limited
point source in the L$'$ AO image.

There is no seeing-limited halo evident in the final image and thus the sensitivity beyond the
first airy ring is determined only by the total amount of integration time covered at that
position. The best sensitivity is obtained in an annulus with 
inner radius just outside the first airy ring at 0.19$'$$'$ to
an outer radius of 4.72$'$$'$ from V488 Per. Adopting the {\it Gaia} EDR3 parallax
(Table \ref{tab:V488Per_param}) results in a distance to V488 Per of 173.5$\pm$0.4\,pc
and separation-space probed by the L$'$ AO image of 33 to 820\,AU.

Interpolating between the {\it WISE} W1 and W2 magnitudes, we obtain an L$'$ magnitude
for V488 Per of $\approx$9. Using V488 Per as a comparison star, we estimate that we are sensitive
to point sources as faint as L$'$$\sim$13~magnitudes at the 5$\sigma$ level
in the best sensitivity areas of the AO image.
Consulting \citet{Leggett_1992} and \citet{Chabrier_2000},
and using the adopted age for the $\alpha$ Per cluster of 80\,Myr (see Section \ref{sec:intro}),
we obtain mass limits of $\sim$0.15$\,M_{\sun}$; we show this mass limit in relation to the
doppler monitoring mass sensitivities in Figure \ref{fig:MassLimit}.
This mass limit is more sensitive than the $\sim$0.6\,M$_{\sun}$ limit obtained
from {\it HST} 1.3\,$\mu$m imaging reported in \citet{patience02}.

We do not assess binarity of V488 Per beyond 800\,AU; \citet{moor21} 
discuss problems with the widely-separated candidate companions identified
in \citet{zuckerman15}.

\subsubsection{\large \bf Stellar Inclination} 
\label{sec:inc}

It is desirable to try and remove the sin$i$ ambiguity in the doppler mass sensitivity limits.
Without any detected companions (nor resolved imaging of the circumstellar disks)
it is not possible to assess the inclination angle of the planetary system around V488 Per.
As a proxy, we will use the inclination angle of the stellar rotation axis, while noting that it
need not be the case that the two axes are aligned.

With a measurement of the rotation period of V488 Per from photometric monitoring and
the rotational broadening of the stellar photospheric absorption lines from APF data
we can obtain the stellar inclination angle. From the UT 2017 Dec 08 APF spectrum
we measure a $v$sin$i$ value of 4.5$\pm$1.5\,km\,s$^{-1}$ following the methodology of
\citet{strassmeir90}.
To obtain the stellar rotation period we analyzed {\it TESS} lightcurve data.
However, the {\it TESS} data reveal only mild variability for V488 Per, not the
clear periodic spot modulation that would be expected for a young, Sun-like star
(Figure \ref{fig:TessLC}).

The result from {\it TESS} stands in stark contrast to variability
measurements for V488 Per($=$AP\,70) in the literature.
Past ground-based observations covering multiple rotational periods clearly detect
rotational modulation of starspots on the surface of V488 Per, finding rotational periods of
123.5~hours (or 5.15~days) in \citet{Stauffer_1985},
6.4$\pm$0.1~days in \citet{allain96},
and 5.98~days in \citet{heinze18}
(ATO\,J052.0779+48.6632, ATLAS ObjID 166390520779906601). 

To place V488 Per in context, we examined {\it TESS} lightcurve data for other
$\alpha$ Per stars with measured
rotational periods in \citet{Stauffer_1985} and \citet{allain96} that had B$-$V colors and
$v$sin$i$ comparable to V488 Per 
(we considered stars with $v$sin$i$$\lesssim$10\,km\,s$^{-1}$ from \citealt{Stauffer_1985}).
A comparison of these stars' {\it TESS} lightcurves to that for V488 Per is shown in Figure
\ref{fig:TessLC}. Each of the additional examined stars shows clear periodic variability
with periods on the order of 4-6~days.

Why does V488 Per not exhibit similar behavior as other $\alpha$ Per stars in 
the {\it TESS} data?
We believe the answer is starspot cycles, and the star AP\,14 shown in Figure \ref{fig:TessLC}
demonstrates the evolution of spot modulation amplitude over the {\it TESS} monitoring
period ($\approx$27~days). In this interpretation, we just happened to catch
V488 Per with {\it TESS} at the minimum of its starspot cycle which had a duration of at least
$\sim$1~month. AP\,14 was captured during its decay phase which probably also took on the order
of at least $\sim$1~month. The other stars shown in Figure \ref{fig:TessLC} were seen
during times of near-maximum activity.

Based on this interpretation, we adopt the average of literature-measured rotational periods for
V488 Per of 5.8~days. To calculate the stellar inclination angle, we further adopt
the stellar radius of 0.76\,R$_{\odot}$ from the spectral energy distribution fit in Figure \ref{fig:SED}.
With these values we calculate a stellar inclination angle of 45$\pm$12$^{\circ}$. This suggests 
that the spectroscopic mass sensitivity values shown in Figure \ref{fig:MassLimit}
should be increased by a factor of $\sim$1.4
{\it if the stellar and planetary system inclination angles are the same}.

\subsection{\large \bf Circumstellar Dust Properties}

While the AO image is not capable of characterizing the dust around V488 Per,
it does effectively eliminate any possibility that the excess emission is due to
a background contaminating object. Any such object would be as bright as V488 Per
in the L$'$-band (Figure \ref{fig:SED}) and would be obvious even at separations interior
to the first airy ring. We conclude with high confidence that all infrared excess emission
seen toward V488 Per is associated with the star.

\subsubsection{\large \bf Dust Architecture}

Figure \ref{fig:SED} collects optical to far-infrared photometric measurements for
V488 Per and fits them with a simplistic model consisting of a stellar photosphere
and two blackbody components.
The {\it Herschel} data clearly confirms that there exists a cooler dust population
in the outer planetary system of V488 Per that was hinted at previously
from {\it WISE} 22.09\,$\mu$m photometry \citep{Zuckerman_2012}.
The blackbody temperature of this outer dust component is reasonably constrained
to $\approx$130\,K while the inner dust population is fit acceptably with a blackbody
having temperature of $\approx$800\,K.

We interpret the two dust populations as separate rings of material within the
V488 Per planetary system. This interpretation is motivated by past results
indicating separate inner and outer planetary system debris disks for other
multiple-blackbody component systems
(e.g., \citealt{backman09}; \citealt{morales09}; \citealt{su09}; \citealt{kennedy14}; \citealt{Vican_2016}).
For the luminosity of V488 Per (calculated with the spectral energy distribution-fit
temperature and radius in L$=$4$\pi$R$^2$T$^4$; see Table \ref{tab:V488Per_param}) 
this would place blackbody-emitting grains having the above temperatures
at orbital separations of $\sim$0.07\,AU and $\sim$2.7\,AU.
The widths of these two putative disks are not well-constrained with the available data
and can be better characterized if the grain composition can be determined (see below)
and/or if resolved imaging can be obtained.

\subsubsection{\large \bf Grain Composition}
\label{sec:dust}

Typically, strong solid-state emission features are seen in the mid-infrared spectra
of exceptionally dusty main sequence stars (e.g., \citealt{Honda_2004}; \citealt{chen06};
\citealt{Lisse_2008};  \citealt{Melis_2010}; \citealt{olofsson12};
\citealt{lisse17}; \citealt{melis21}).
V488 Per is thus unusual in not showing any obvious features in its COMICS
N-band spectrum (Figures \ref{fig:compare} and \ref{fig:dust}). 
Figure \ref{fig:compare} compares the COMICS mid-infrared spectrum of V488 Per
with {\it Spitzer} IRS spectra of exemplar extreme debris disk systems having well-detected
solid-state emission features.
To date, stars having fractional infrared 
luminosities $\gtrsim$1\% all have strong solid-state emission features 
present in their mid-infrared
spectra that can be ruled out in the spectrum of V488 Per
(several new extremely dusty main sequence stars have yet to be observed
with mid-infrared spectrometers; e.g., those stars from \citealt{gaidos19},
\citealt{tajiri20}, and \citealt{moor21}). This stands for all types of dust grain
species currently known, whether they be silica-rich (e.g., \citealt{chen06}; 
\citealt{rhee08};  
\citealt{Melis_2010}; 
\citealt{fujiwara12dust}), 
dominated by amorphous and/or crystalline silicates (e.g., \citealt{honda04}; 
\citealt{song05}; 
\citealt{chen06}; 
\citealt{rhee07}; 
\citealt{olofsson12}; 
\citealt{melis21}), 
or stars with yet-to-be-identified emission features (e.g., \citealt{melis13}; 
\citealt{lisse17}). 
When compared against less dusty stars (fractional infrared luminosities $\sim$0.01\%),
it is possible that the noise in the COMICS spectrum of V488 Per could hide weaker
solid-state emission features. This is evidenced by comparison to HD\,69830 (e.g., \citealt{beichman05})
in Figure \ref{fig:compare},
and is similarly true for other such systems (e.g., \citealt{chen06}; 
\citealt{moor09}; 
\citealt{mittal15}). 
While it is possible that some weaker solid state emission features could be hidden
in the COMICS spectrum presented for V488 Per, it remains the case that it is unusual when 
compared against other stars having huge amounts of inner planetary system
dust ($\tau$$\gtrsim$1\%).

We discuss some possible explanations for why no strong solid state emission feature
is seen in the COMICS spectrum of V488 Per and elaborate on them further here and
in Section \ref{sec:conclusions}.
To generate the observed featureless mid-infrared spectrum, either 
1) only a weak solid-state emission feature is present as discussed above;
2) dust grains are larger than $\sim$10\,$\mu$m in size, at which point
solid-state emission features near a wavelength of 10\,$\mu$m become severely
muted (e.g., Figure \ref{fig:dust}; see also \citealt{min07} and Section 4.3 of \citealt{olofsson12}); 
and/or 3) the dust composition could be dominated by species which do not produce obvious
mid-infrared solid-state emission features, like amorphous carbon or metallic iron.
Possibility 2) seems highly unlikely given that the observed dust is almost certainly produced
via collisions of rocky bodies which should eventually be ground down to radiation blow-out sizes
($\lesssim$0.1\,$\mu$m for V488 Per; e.g., \citealt{artymowicz88})
before being removed from the planetary system.

To explore possibility 1) in more detail, we constructed solid-state emission feature models
for V488 Per that would be consistent with the observed COMICS spectrum.
In these models, we utilize a blackbody component for the disk continuum emission and 
an emission feature component computed from the linear combination of dust absorption 
coefficients multiplied by a blackbody; the solid angle of each component is adjusted to find
agreement with the COMICS spectrum.
We use dust absorption coefficients from \citet{min07} for silicates and silica grain
species. Since emission from amorphous carbon or
metallic iron is indistinguishable from continuum emission in the 8-13\,$\mu$m wavelength 
region (e.g., Figure 2 of \citealt{chen06}; Figure 3 of \citealt{lisse08}), 
we do not explore those grain species and cannot constrain their presence.
Figure \ref{fig:dust} shows example dust emission models that are consistent with
the COMICS data. There are a wide range of models that could be consistent with the data,
but all such models have a total solid-state emitting dust grain
mass $\lesssim$10$^{20}$\,g (a couple times the mass of the asteroid Ida).
In comparison, other extremely dusty main sequence stars ($\tau$$\gtrsim$1\%)
have been found to have
masses for their solid-state emitting grains in the range of 10$^{21}$-10$^{23}$\,g
(e.g., \citealt{chen06}; \citealt{lisse08}; \citealt{lisse09}).
Systems like HD\,69830 ($\tau$$\sim$0.01\% and
whose mid-infrared solid-state emission features could
hide in the V488 Per COMICS spectrum) have solid-state emitting grain 
dust masses that encompass
the limits obtained for the models shown in Figure \ref{fig:dust}, 10$^{19}$-10$^{21}$\,g
(e.g., \citealt{beichman05}; \citealt{chen06}; \citealt{lisse07,lisse12}).

If V488 Per is like other extremely dusty main sequence stars and hosts a small grain
population with mass of $\gtrsim$10$^{21}$\,g, then these small grains must not produce
strong solid-state emission features in the 8-13\,$\mu$m range. This limits the contribution
from typically-seen silicates and silica to $<$10$^{20}$\,g and requires an
order of magnitude more mass of dust from species like amorphous carbon and
metallic iron. Given that inner planetary systems (including our own) are found to be
deficient in carbon (e.g., \citealt{jura06} and references therein), we give preference
to a scenario where these dust grains are predominantly metallic iron.

\section{\large \bf Conclusions}
\label{sec:conclusions}

We present an investigation into the overall architecture of the V488 Per planetary system; 
Figure \ref{fig:MassLimit} shows companion sensitivity limits, while
Figures \ref{fig:SED} and \ref{fig:dust} summarize dust properties. 

Blackbody fits to infrared excess measurements indicate two dust populations
having temperatures of 800 and 130\,K, or orbital separations of roughly 0.07 and 2.7\,AU
respectively. 
To constrain the true extent of these dust populations within the V488 Per planetary system
will require additional data. Longer wavelength observations (e.g., with ALMA) can determine
the outermost extent of the cool dust population while optical/infrared interferometric observations
might be capable of assessing the presence of very close-in dust.

Assuming a separated two-dust-belt interpretation is appropriate, we limit any companions
within the gap between the belts to have mass $\lesssim$10\,M$_{Jup}$.
This limit could be improved with precision radial velocity monitoring over a few year duration.
Stellar-mass companions are effectively ruled out for most separations $<$800\,AU from
V488 Per. Deeper high spatial resolution imaging (PSF full-width at half-max 
$\lesssim$50\,mas) or sparse-aperture masking interferometry could push to sub-stellar
mass limits in this region.

The lack of any obvious solid-state emission features in
the COMICS mid-infrared spectrum could be suggestive of dust
particles composed primarily of metallic iron. If true, this might indicate 
the erosive bombardment and subsequent
grinding down of ejecta from something resembling the deep interior of a differentiated planet.
\citet{clement21} discuss how such a scenario may have played out during the formation
of Mercury in the solar system. It is more likely to obtain such a scenario in exoplanetary systems
with inner planetary systems tightly packed with Super-Earth/Earth-type
planets (e.g., \citealt{pu15}; \citealt{izidoro19}; \citealt{clement21}, and references therein),
and as such it is not unreasonable to suggest (based on the available data)
that V488 Per could potentially be in the process of forming a Mercury-like planet.
Planned GTO {\it JWST}/MIRI observations
of V488 Per will be valuable in assessing the presence and variability of any solid-state
emission features and thus further exploring this hypothesis.

\acknowledgments

C.M.\ acknowledges support from NASA ADAP grant 18-ADAP18-0233.
B.K.\ acknowledges support from the APS M.\ Hildred Blewett Fellowship.
AWH acknowledges NSF grant AST-1517655.
We thank the anonymous referee for useful comments that helped improve this paper.
The authors wish to thank Bradford Holden for assistance in scheduling the APF observations presented in this paper. We thank George Rieke for useful discussion.
Research at Lick Observatory is partially supported by a generous gift from Google.
Some of the data presented herein were obtained at the W.M.\ Keck Observatory, which is operated as a scientific partnership among the California Institute of Technology, the University of California and the National Aeronautics and Space Administration. The Observatory was made possible by the generous financial support of the W.M.\ Keck Foundation.
The authors wish to recognize and acknowledge the very significant cultural role and reverence that the summit of Maunakea has always had within the indigenous Hawaiian community.  We are most fortunate to have the opportunity to conduct observations from this mountain.
Similarly, we acknowledge that Lick Observatory resides on land traditionally inhabited by the 
Muwekma Ohlone Tribe of Native Americans.
This paper includes data collected by the {\it TESS} mission, which are publicly available from the Mikulski Archive for Space Telescopes (MAST).
Funding for the {\it TESS} mission is provided by NASA's Science Mission directorate.
This work is based in part on observations made with {\it Herschel}, 
an ESA space observatory with science instruments provided by 
European-led Principal Investigator consortia and with important participation from NASA.
Support for {\it Herschel} work was provided by NASA through an award issued by JPL/Caltech.
This research has made use of NASA's Astrophysics Data System, the SIMBAD database,
and the VizieR service.

\facilities{Subaru(COMICS), APF(Levy), Shane(Hamilton), Keck(NIRC2), {\it Herschel}(PACS), {\it TESS}}

}

\clearpage

\clearpage

\begin{figure}
 \centering
 \begin{minipage}[!t]{87mm}
  \includegraphics[width=90mm]{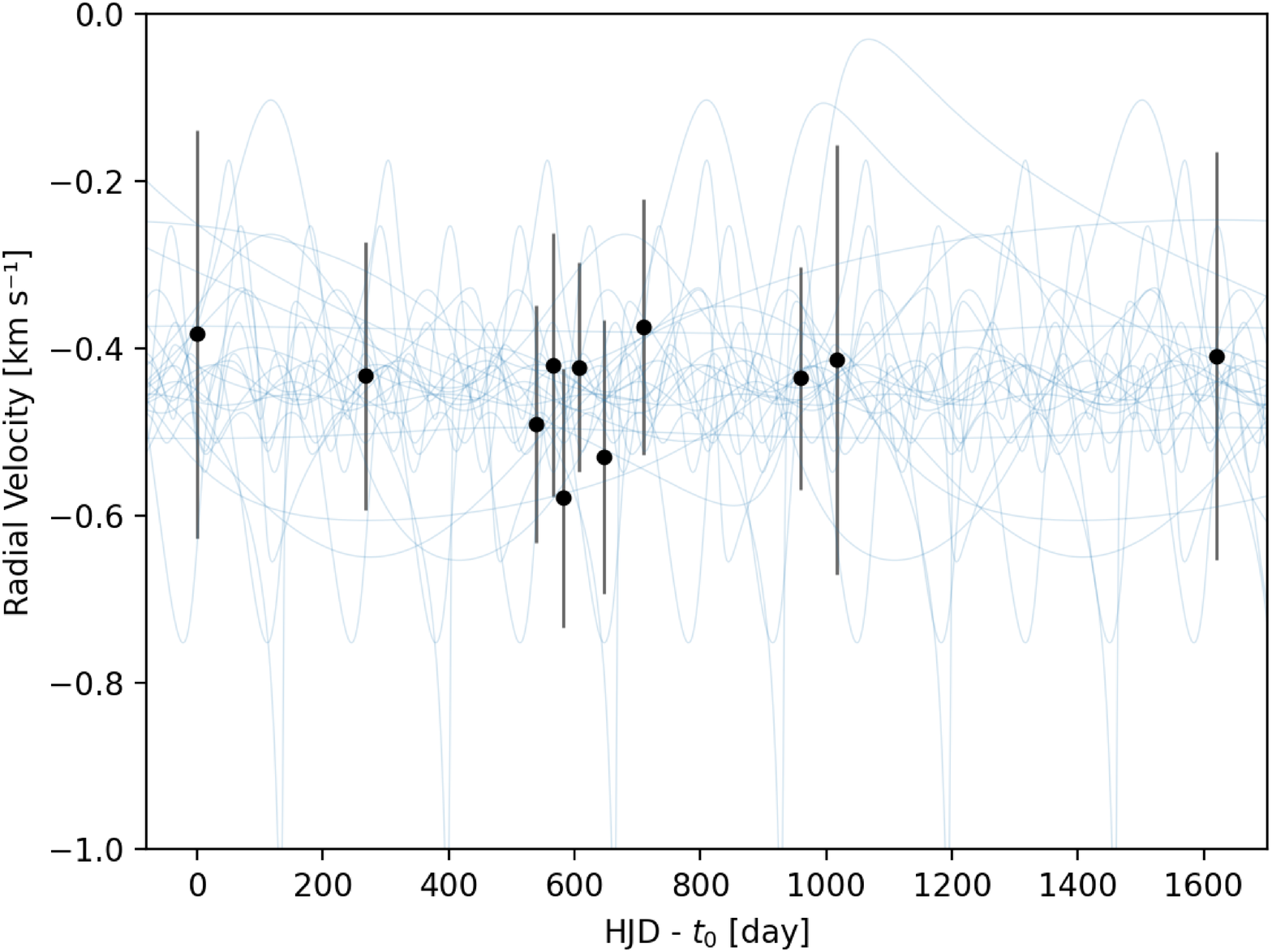}
 \end{minipage}
 \begin{minipage}[!t]{73mm}
  \includegraphics[width=67.5mm]{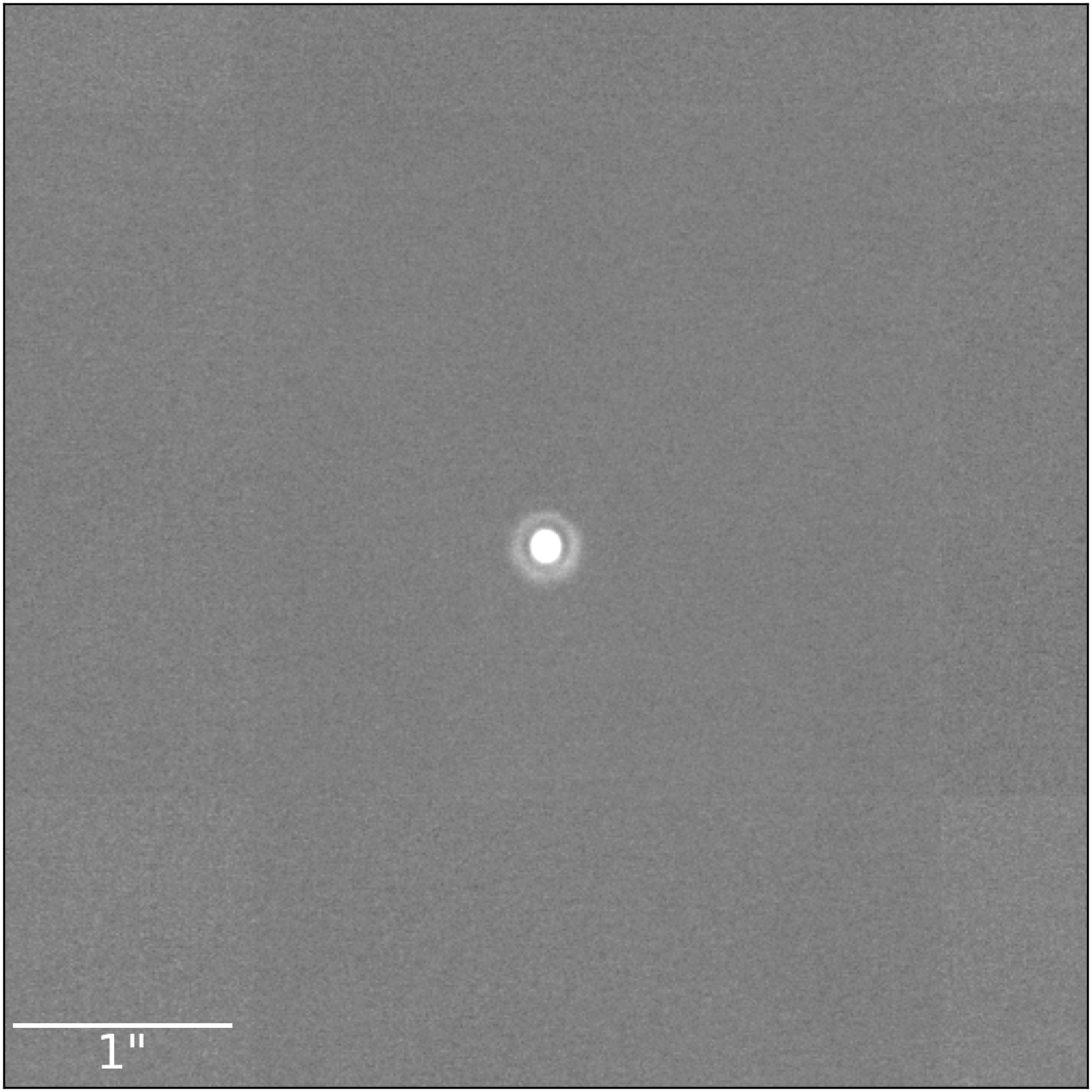} 
 \end{minipage}
 \\*
 \begin{minipage}[!t]{160mm}
  \includegraphics[width=155mm]{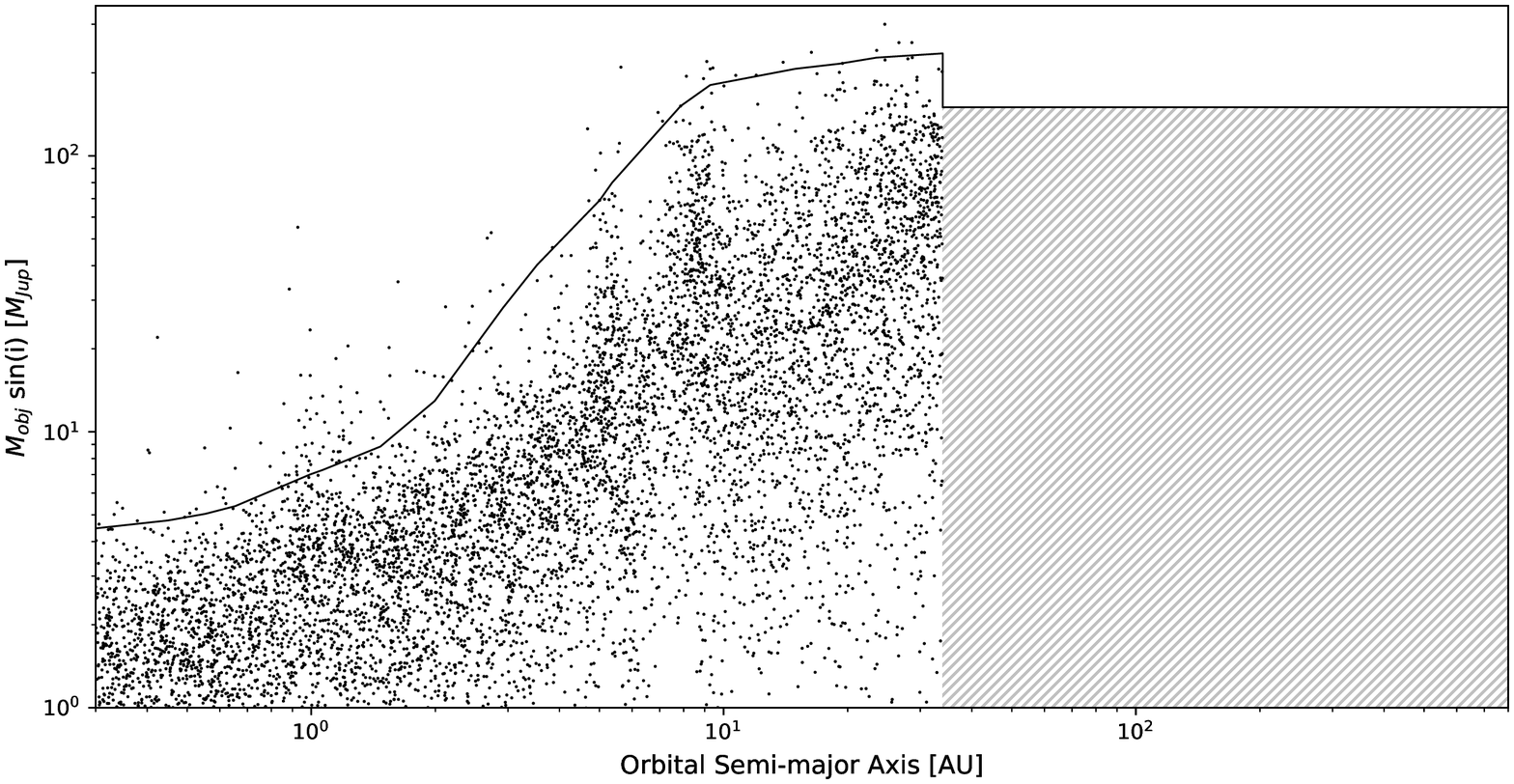} 
 \end{minipage}
 \caption{\label{fig:MassLimit} \large{Radial velocity monitoring
               and adaptive optics imaging results and mass sensitivity limits derived from them. 
               {\it Top Left Panel:} Time series of radial velocity measurements for 
               V488 Per (black dots with error bars) 
               with a selection of representative {\sf The Joker} allowable solutions overplotted
               (blue dotted curves).
               t$_{0}$ is HJD 2457448.75332 (UT 2016 Mar 01), the date of the first observation of 
               V488 Per at Lick Observatory. The median radial velocity for all measurements shown is 
               $-$0.423\,km\,s$^{-1}$.
               {\it Top Right Panel:} Keck AO L$'$ image of V488 Per; the first 
               airy ring is clearly visible around the star. 
               North is up and East is left, no other sources are detected in the imaged field of view.
               {\it Bottom Panel:} Mass sensitivity limits obtained as described in Section \ref{sec:disc};
               masses are $m$sin$i$ only for the spectroscopic curve covering semi-major axes $<$30\,AU,
               there is no inclination angle ambiguity for the AO imaging mass sensitivity line
               shown for semi-major axes $>$30\,AU.
               Dots are from allowable solutions obtained with {\sf The Joker}, the hatched
               region shows companions that would not have been detected with AO imaging;
               above the solid curve we expect we would have detected any such companions if they
               were present.}}
\end{figure}

\clearpage

\begin{figure*}
 \centering
 \includegraphics[width=180mm]{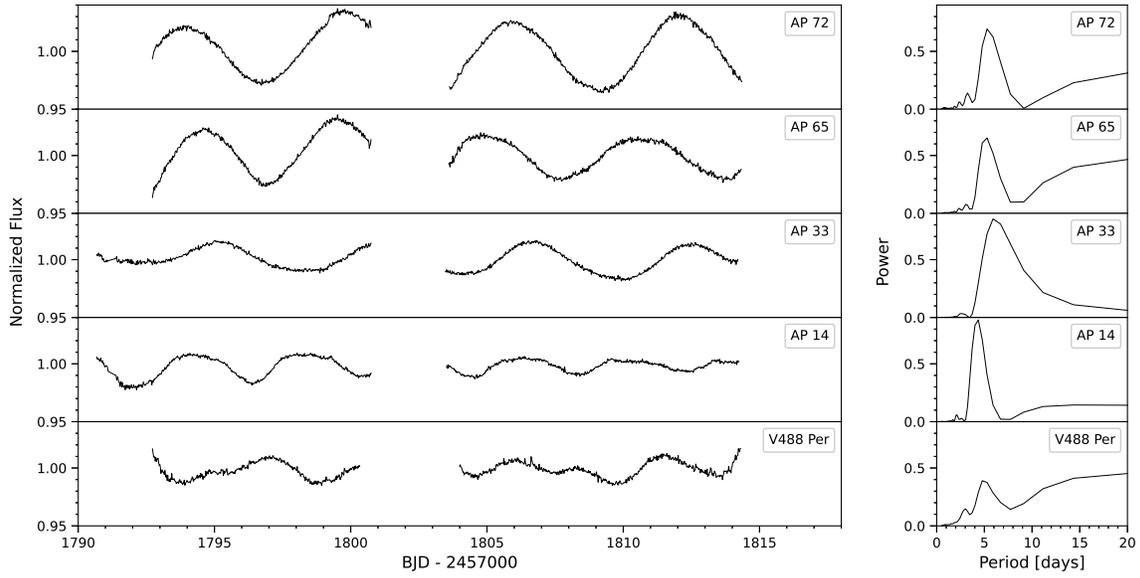}
 \caption{\label{fig:TessLC} \large{TESS light curves of V488 Per and other 
               $\alpha$ Per stars having similar B$-$V colors and $v$sin$i$ as V488 Per.
               The right column shows periodograms calculated from these lightcurves;
               a marginal period around 5 days is possible for V488 Per while clear
               periods of 4-6~days are found for the other stars. Of note is AP\,14, which
               shows a reduction in the amplitude of its periodic signal over the {\it TESS}
               monitoring period and ends with variability levels comparable to V488 Per.}}
\end{figure*}

\clearpage

\begin{figure}
 \centering
  \includegraphics[width=160mm]{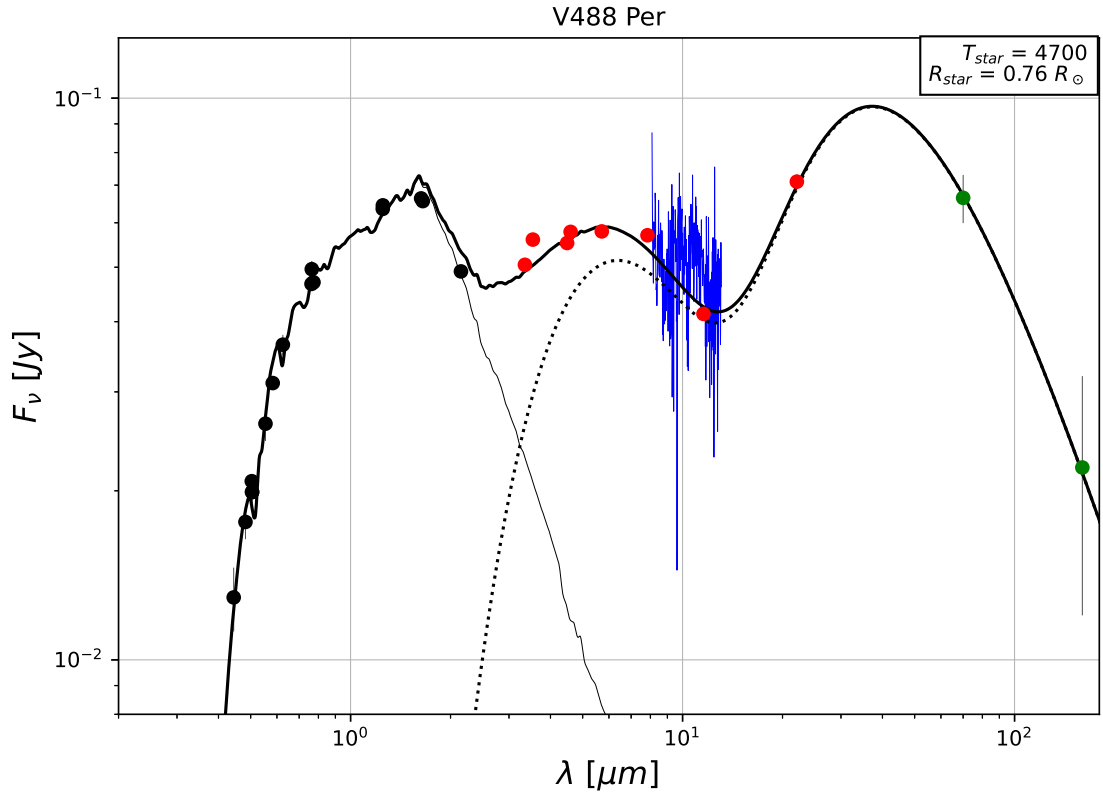}
 \caption{\label{fig:SED} \large{Spectral energy distribution including the COMICS mid-infrared
               spectrum of V488 Per.
               The thin black curve connecting optical and near-infrared data points
               is a synthetic stellar atmospheric spectrum with temperature and radius as
               indicated on the figure panel. 
               The dotted curve is a sum of two blackbodies having temperatures of
               800 and 130\,K. The thick black curve is the sum of all the individual model components.
               Data points used to inform the stellar photospheric model
               are AAVSO \citep{henden15}, {\it Gaia} photometry 
               \citep{gaia20} and 2MASS magnitudes \citep{skrutskie06}.
               Excess measurements are from {\it WISE} \citep{cutri12}, 
               IRAC from \citet{Zuckerman_2012},
               and {\it Herschel} (in green, see text). 
               The blue curve between the IRAC 7.8\,$\mu$m and {\it WISE}
               11.56\,$\mu$m measurements is the COMICS N-band spectrum.
               Vertical lines in data points indicate the measurement
               uncertainty. Some measurement uncertainties are smaller
               than the point sizes on the plot.
               }}
\end{figure}

\clearpage

\begin{figure}
 \centering
 \begin{minipage}[!t]{110mm}
  \includegraphics[width=110mm]{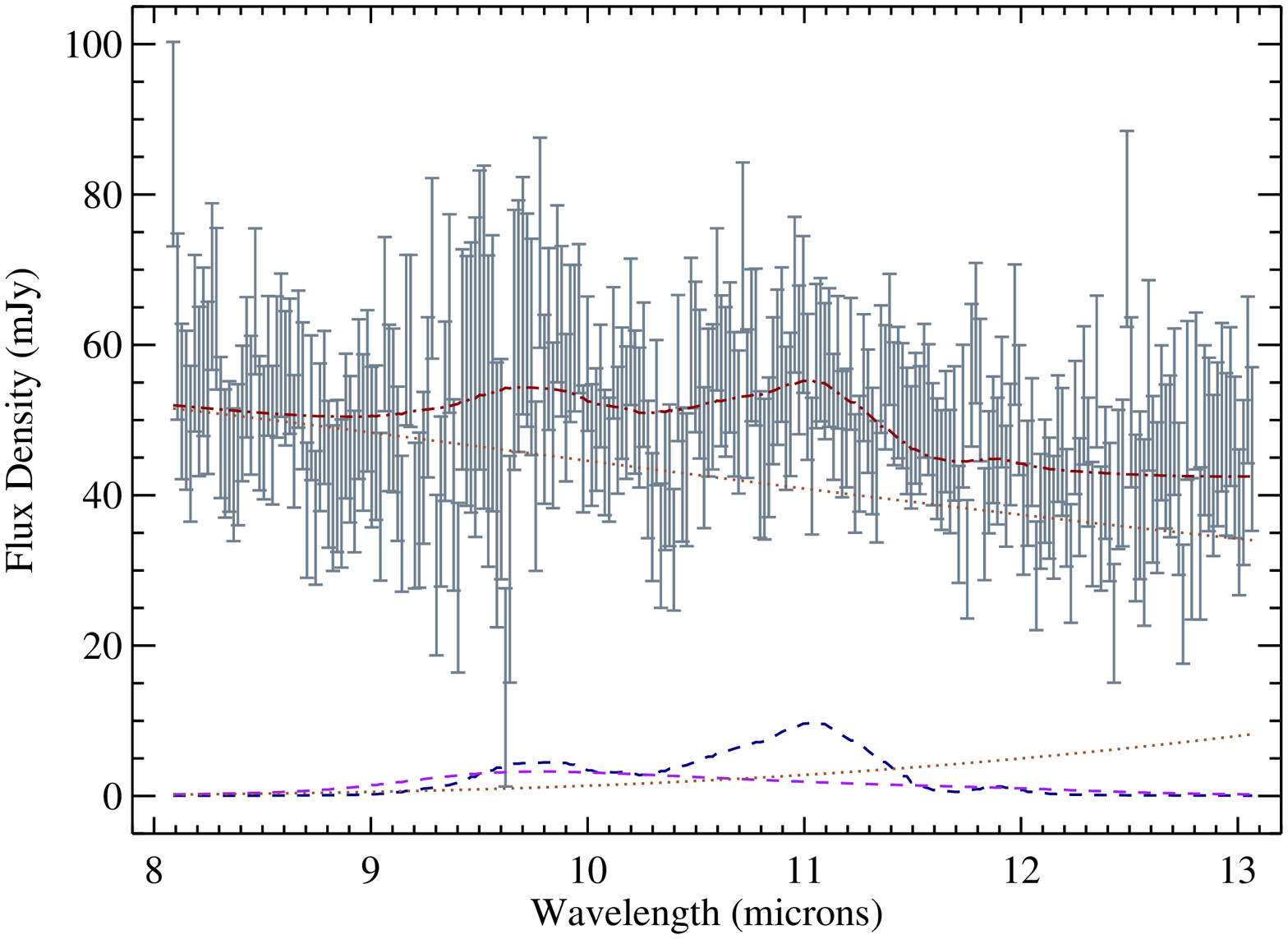} 
 \end{minipage}
 \\
 \begin{minipage}[!t]{110mm}
  \includegraphics[width=110mm]{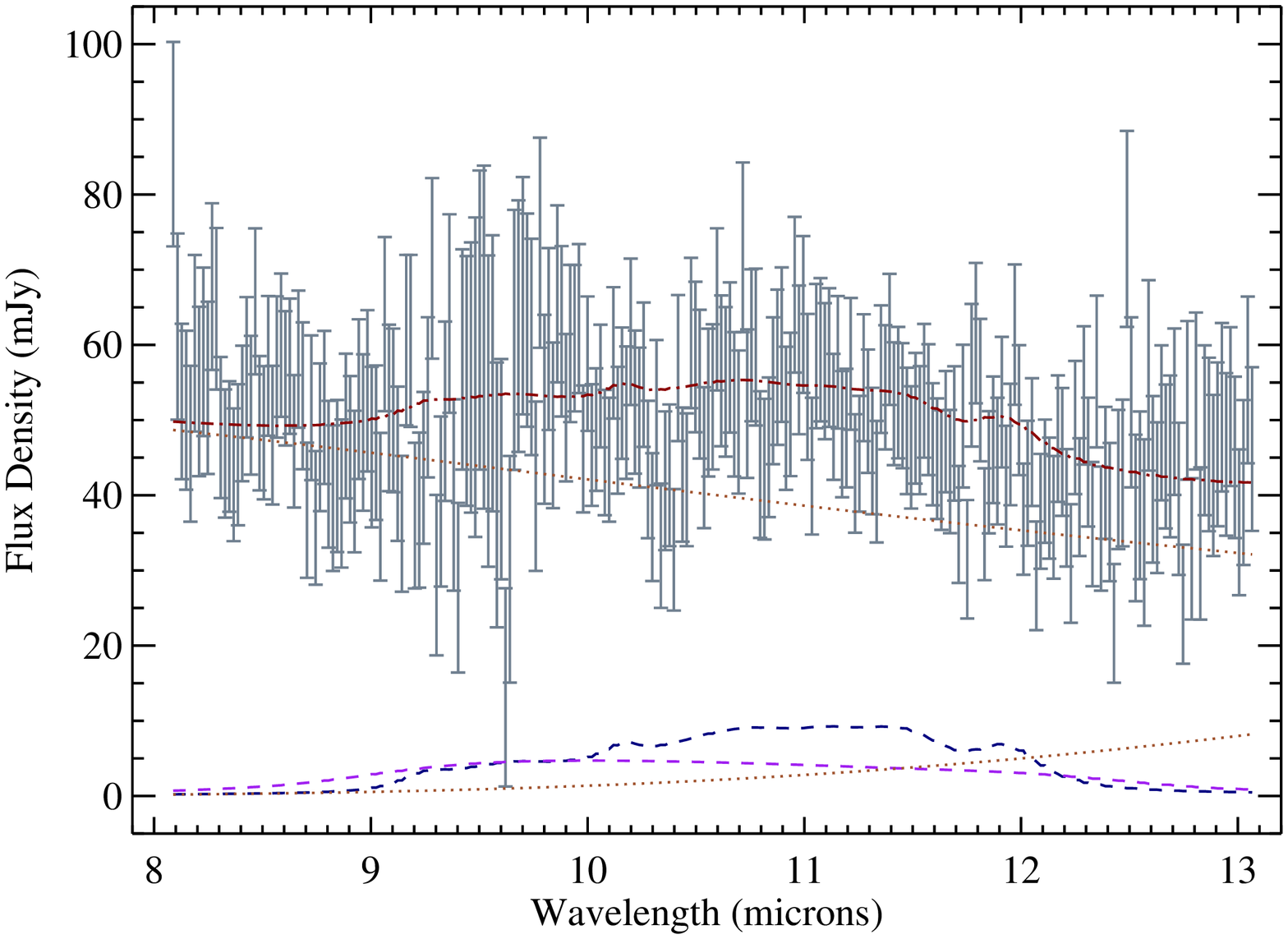} 
 \end{minipage}
 \caption{\label{fig:dust} \large{Example solid-state emission models that are consistent with
               the V488 Per COMICS spectrum. In each panel, the V488 Per spectrum
               is shown in grey (spectral samples and associated uncertainties),
               orange dotted curves are 800 and 130\,K blackbodies,
               purple and blue dashed curves are amorphous and crystalline silicates
               respectively, and the red dot-dashed curve is the sum of all components shown.    
               We do not include amorphous carbon or metallic iron in these models as their emission
               is indistinguishable from that of the continuum components; it is not possible to constrain
               amorphous carbon or metallic iron with the available data (see Section \ref{sec:dust}).           
               {\it Top Panel:} Small dust grain emission model where the amorphous and crystalline silicates
               have a size of 0.1\,$\mu$m. This model has a total mass of solid-state emitting grains of
               $\approx$4$\times$10$^{19}$\,g, or the mass of the asteroid Ida.
               While this model is a reasonable match to the data, we caution that the various
               wiggles seen in the spectrum could instead be related to its low signal-to-noise ratio; e.g.,
               such wiggles could be from low-level, low-frequency noise in the detector electronics.
               {\it Bottom Panel:} Large dust grain emission model where the amorphous and crystalline
               silicates have a size of 4.0\,$\mu$m. The muting effect of the larger grain sizes on
               the solid-state emission features is evident. For grain sizes $>$5\,$\mu$m
               no clear peaks from
               the solid-state transitions are present and it is difficult to distinguish them from
               the continuum (see discussion in Section \ref{sec:dust}). This model has a total mass of 
               solid-state emitting grains that is a few times the mass of Ida.}}
\end{figure}

\clearpage



\begin{table}[h!]
\caption{{\large V488 Per Stellar Parameters} \label{tab:V488Per_param}}
\begin{center}
\begin{tabular}{lcc}
\hline
\hline
Parameter  & Value &  Reference   \\
\hline
V$_{\rm mag}$                & 12.89$\pm$0.05     & \citealt{henden15} \\
$\pi$ [mas]                      &   5.763$\pm$0.012 & \citealt{gaia20}  \\
T$_{eff}$ [K]                    & 4700                       & This paper, Figure \ref{fig:SED} \\
$R_{\star}$ [$R_{\sun}$] & 0.76                        & This paper, Figure \ref{fig:SED} \\
$L_{\star}$ [$L_{\sun}$]  & 0.25                        & This paper \\
$M_{\star}$ [$M_{\sun}$] & 0.84                       & \citealt{wright11}; \citealt{stassun19} \\
$i$ [$^\circ$]                    &   45$\pm$12          & This paper, Section \ref{sec:inc} \\
\hline
\end{tabular}
\end{center}
\end{table}

\begin{table*}[ht]
\caption{{\large Lick Observatory Radial Velocity Measurements} \label{tab:obs}}
\begin{center}
\begin{tabular}{ccccc}
\hline 
\hline
UT Date  & HJD & Instrument  & SNR$^{a}$  & V$_{r}$ \\
		&		  & 		&	&  km\,s$^{-1}$ \\
\hline
2016 Mar 01 & 2457448.75332 & APF & 	10.14 & --0.383 $\pm$ 0.244  \\
2016 Nov 01 & 2457716.90075 & APF  &  18.03 & --0.433 $\pm$ 0.160 \\
2017 Aug 21 & 2457987.00913 & APF  &  19.81 & --0.578 $\pm$ 0.155 \\
2017 Sep 18 & 2458014.75714 & APF  &  21.09 & --0.420 $\pm$ 0.158 \\
2017 Oct 04 & 2458030.81506 & APF  &  19.66 & --0.530 $\pm$ 0.164 \\
2017 Oct 30 & 2458056.78022 & APF  &  21.20 & --0.423 $\pm$ 0.158 \\
2017 Dec 08 & 2458095.88214 & APF  &  20.91 & --0.490 $\pm$ 0.142 \\
2018 Feb 08 & 2458157.68050 & APF  &  19.31 & --0.375 $\pm$ 0.142 \\
2018 Oct 17 & 2458408.78883 & APF  &  22.19 & --0.436 $\pm$ 0.113 \\
2018 Dec 13 & 2458465.72084 & APF  &  11.68 & --0.410 $\pm$ 0.257 \\
2020 Aug 07 & 2459068.99811 & Hamilton  &  18.60 & --0.409 $\pm$ 0.244 \\
\hline
\end{tabular}
\\
$^{a}$The signal-to-noise ratio per pixel, measured near 6475\,\AA .
\end{center}
\label{table:radvelobs}
\end{table*}


\begin{thebibliography}{83}
\expandafter\ifx\csname natexlab\endcsname\relax\def\natexlab#1{#1}\fi

\bibitem[\protect\astroncite{{Absil} {\em et~al.\/}}{2021}]{absil21}
{Absil}, O., {\em et~al.\/} 2021, {\em arXiv e-prints\/}, arXiv:2104.14216

\bibitem[\protect\astroncite{{Allain} {\em et~al.\/}}{1996}]{allain96}
{Allain}, S., {Fernandez}, M., {Martin}, E.~L., \& {Bouvier}, J. 1996, {\em
  \aap\/}, {\bf 314}, 173

\bibitem[\protect\astroncite{{Artymowicz}}{1988}]{artymowicz88}
{Artymowicz}, P. 1988, {\em \apjl\/}, {\bf 335}, L79

\bibitem[\protect\astroncite{{Backman} {\em et~al.\/}}{2009}]{backman09}
{Backman}, D., {\em et~al.\/} 2009, {\em \apj\/}, {\bf 690}, 1522

\bibitem[\protect\astroncite{{Ballering} {\em et~al.\/}}{2014}]{ballering14}
{Ballering}, N.~P., {Rieke}, G.~H., \& {G{\'a}sp{\'a}r}, A. 2014, {\em \apj\/},
  {\bf 793}, 57

\bibitem[\protect\astroncite{{Beichman} {\em et~al.\/}}{2005}]{beichman05}
{Beichman}, C.~A., {\em et~al.\/} 2005, {\em \apj\/}, {\bf 626}, 1061

\bibitem[\protect\astroncite{{Bonsor} {\em et~al.\/}}{2013}]{bonsor13}
{Bonsor}, A., {Raymond}, S.~N., \& {Augereau}, J.-C. 2013, {\em \mnras\/}, {\bf
  433}, 2938

\bibitem[\protect\astroncite{{Bonsor} {\em et~al.\/}}{2014}]{bonsor14}
{Bonsor}, A., {Raymond}, S.~N., {Augereau}, J.-C., \& {Ormel}, C.~W. 2014, {\em
  \mnras\/}, {\bf 441}, 2380

\bibitem[\protect\astroncite{Chabrier {\em et~al.\/}}{2000}]{Chabrier_2000}
Chabrier, G., Baraffe, I., Allard, F., \& Hauschildt, P. 2000, {\em The
  Astrophysical Journal\/}, {\bf 542}, 464

\bibitem[\protect\astroncite{{Chen} {\em et~al.\/}}{2006}]{chen06}
{Chen}, C.~H., {\em et~al.\/} 2006, {\em \apjs\/}, {\bf 166}, 351

\bibitem[\protect\astroncite{{Clement} {\em et~al.\/}}{2021}]{clement21}
{Clement}, M.~S., {Chambers}, J.~E., \& {Jackson}, A.~P. 2021, {\em \aj\/},
  {\bf 161}, 240

\bibitem[\protect\astroncite{{Currie} {\em et~al.\/}}{2011}]{currie11}
{Currie}, T., {Lisse}, C.~M., {Sicilia-Aguilar}, A., {Rieke}, G.~H., \& {Su},
  K. Y.~L. 2011, {\em \apj\/}, {\bf 734}, 115

\bibitem[\protect\astroncite{{Cutri} \& {et al.}}{2012}]{cutri12}
{Cutri}, R.~M. \& {et al.} 2012, {\em VizieR Online Data Catalog\/}, II/311

\bibitem[\protect\astroncite{{Fujiwara} {\em et~al.\/}}{2009}]{fujiwara09}
{Fujiwara}, H., {\em et~al.\/} 2009, {\em \apjl\/}, {\bf 695}, L88

\bibitem[\protect\astroncite{{Fujiwara} {\em
  et~al.\/}}{2012{\natexlab{a}}}]{fujiwara12dust}
--- 2012{\natexlab{a}}, {\em \apjl\/}, {\bf 749}, L29

\bibitem[\protect\astroncite{{Fujiwara} {\em
  et~al.\/}}{2012{\natexlab{b}}}]{fujiwara12}
--- 2012{\natexlab{b}}, {\em \apjl\/}, {\bf 759}, L18

\bibitem[\protect\astroncite{{Gaia Collaboration} {\em
  et~al.\/}}{2020}]{gaia20}
{Gaia Collaboration}, {Brown}, A.~G.~A., {Vallenari}, A., {Prusti}, T., {de
  Bruijne}, J.~H.~J., {Babusiaux}, C., \& {Biermann}, M. 2020, {\em arXiv
  e-prints\/}, arXiv:2012.01533

\bibitem[\protect\astroncite{{Gaidos} {\em et~al.\/}}{2019}]{gaidos19}
{Gaidos}, E., {\em et~al.\/} 2019, {\em \mnras\/}, {\bf 488}, 4465

\bibitem[\protect\astroncite{{Gomes} {\em et~al.\/}}{2005}]{gomes05}
{Gomes}, R., {Levison}, H.~F., {Tsiganis}, K., \& {Morbidelli}, A. 2005, {\em
  \nat\/}, {\bf 435}, 466

\bibitem[\protect\astroncite{{Heinze} {\em et~al.\/}}{2018}]{heinze18}
{Heinze}, A.~N., {\em et~al.\/} 2018, {\em \aj\/}, {\bf 156}, 241

\bibitem[\protect\astroncite{{Henden} {\em et~al.\/}}{2015}]{henden15}
{Henden}, A.~A., {Levine}, S., {Terrell}, D., \& {Welch}, D.~L. 2015, in {\em
  American Astronomical Society Meeting Abstracts \#225\/}, vol. 225 of {\em
  American Astronomical Society Meeting Abstracts\/},  336.16

\bibitem[\protect\astroncite{{Herschel Point Source Catalogue Working Group}
  {\em et~al.\/}}{2020}]{herschel20}
{Herschel Point Source Catalogue Working Group}, {\em et~al.\/} 2020, {\em
  VizieR Online Data Catalog\/}, VIII/106

\bibitem[\protect\astroncite{{Honda} {\em et~al.\/}}{2004}]{honda04}
{Honda}, M., {\em et~al.\/} 2004, {\em \apjl\/}, {\bf 610}, L49

\bibitem[\protect\astroncite{Honda {\em et~al.\/}}{2004}]{Honda_2004}
Honda, M., {\em et~al.\/} 2004, {\em The Astrophysical Journal Letters\/}, {\bf
  610}, L49

\bibitem[\protect\astroncite{Huang {\em
  et~al.\/}}{2020{\natexlab{a}}}]{Huang_2020_a}
Huang, C.~X., {\em et~al.\/} 2020{\natexlab{a}}, {\em Research Notes of the
  AAS\/}, {\bf 4}, 204

\bibitem[\protect\astroncite{Huang {\em
  et~al.\/}}{2020{\natexlab{b}}}]{Huang_2020_b}
--- 2020{\natexlab{b}}, {\em Research Notes of the AAS\/}, {\bf 4}, 206

\bibitem[\protect\astroncite{{Izidoro} {\em et~al.\/}}{2019}]{izidoro19}
{Izidoro}, A., {Bitsch}, B., {Raymond}, S.~N., {Johansen}, A., {Morbidelli},
  A., {Lambrechts}, M., \& {Jacobson}, S.~A. 2019, {\em arXiv e-prints\/},
  arXiv:1902.08772

\bibitem[\protect\astroncite{{Johnson} {\em et~al.\/}}{2012}]{johnson12}
{Johnson}, B.~C., {\em et~al.\/} 2012, {\em \apj\/}, {\bf 761}, 45

\bibitem[\protect\astroncite{{Jura}}{2006}]{jura06}
{Jura}, M. 2006, {\em \apj\/}, {\bf 653}, 613

\bibitem[\protect\astroncite{{Kataza} {\em et~al.\/}}{2000}]{kataza00}
{Kataza}, H., {\em et~al.\/} 2000, in {\em Optical and IR Telescope
  Instrumentation and Detectors\/}, edited by M.~{Iye} \& A.~F. {Moorwood},
  vol. 4008 of {\em Society of Photo-Optical Instrumentation Engineers (SPIE)
  Conference Series\/},  1144--1152

\bibitem[\protect\astroncite{{Kennedy} \& {Wyatt}}{2014}]{kennedy14}
{Kennedy}, G.~M. \& {Wyatt}, M.~C. 2014, {\em \mnras\/}, {\bf 444}, 3164

\bibitem[\protect\astroncite{Leggett}{1992}]{Leggett_1992}
Leggett, S. 1992, {\em The Astrophysical Journal Supplement Series\/}, {\bf
  82}, 351

\bibitem[\protect\astroncite{Lisse {\em et~al.\/}}{2008}]{Lisse_2008}
Lisse, C., Chen, C., Wyatt, M., \& Morlok, A. 2008, {\em The Astrophysical
  Journal\/}, {\bf 673}, 1106

\bibitem[\protect\astroncite{{Lisse} {\em et~al.\/}}{2007}]{lisse07}
{Lisse}, C.~M., {Beichman}, C.~A., {Bryden}, G., \& {Wyatt}, M.~C. 2007, {\em
  \apj\/}, {\bf 658}, 584

\bibitem[\protect\astroncite{{Lisse} {\em et~al.\/}}{2008}]{lisse08}
{Lisse}, C.~M., {Chen}, C.~H., {Wyatt}, M.~C., \& {Morlok}, A. 2008, {\em
  \apj\/}, {\bf 673}, 1106

\bibitem[\protect\astroncite{{Lisse} {\em et~al.\/}}{2009}]{lisse09}
{Lisse}, C.~M., {Chen}, C.~H., {Wyatt}, M.~C., {Morlok}, A., {Song}, I.,
  {Bryden}, G., \& {Sheehan}, P. 2009, {\em \apj\/}, {\bf 701}, 2019

\bibitem[\protect\astroncite{{Lisse} {\em et~al.\/}}{2012}]{lisse12}
{Lisse}, C.~M., {\em et~al.\/} 2012, {\em \apj\/}, {\bf 747}, 93

\bibitem[\protect\astroncite{{Lisse} {\em et~al.\/}}{2017}]{lisse17}
--- 2017, {\em \apjl\/}, {\bf 840}, L20

\bibitem[\protect\astroncite{{Lisse} {\em et~al.\/}}{2020}]{lisse20}
--- 2020, {\em \apj\/}, {\bf 894}, 116

\bibitem[\protect\astroncite{{Melis}}{2016}]{melis16}
{Melis}, C. 2016, in {\em Young Stars \& Planets Near the Sun\/}, edited by
  J.~H. {Kastner}, B.~{Stelzer}, \& S.~A. {Metchev}, vol. 314,  241--246

\bibitem[\protect\astroncite{{Melis} {\em et~al.\/}}{2021}]{melis21}
{Melis}, C., {Olofsson}, J., {Song}, I., {Sarkis}, P., {Weinberger}, A.~J.,
  {Kennedy}, G., \& {Krumpe}, M. 2021, {\em arXiv e-prints\/}, arXiv:2104.06448

\bibitem[\protect\astroncite{{Melis} {\em et~al.\/}}{2010}]{Melis_2010}
{Melis}, C., {Zuckerman}, B., {Rhee}, J.~H., \& {Song}, I. 2010, {\em \apjl\/},
  {\bf 717}, L57

\bibitem[\protect\astroncite{{Melis} {\em et~al.\/}}{2013}]{melis13}
{Melis}, C., {Zuckerman}, B., {Rhee}, J.~H., {Song}, I., {Murphy}, S.~J., \&
  {Bessell}, M.~S. 2013, {\em \apj\/}, {\bf 778}, 12

\bibitem[\protect\astroncite{{Meng} {\em et~al.\/}}{2014}]{meng14}
{Meng}, H. Y.~A., {\em et~al.\/} 2014, {\em Science\/}, {\bf 345}, 1032

\bibitem[\protect\astroncite{Mermilliod {\em et~al.\/}}{2008}]{Mermilliod_2008}
Mermilliod, J.-C., Queloz, D., \& Mayor, M. 2008, {\em Astronomy \&
  Astrophysics\/}, {\bf 488}, 409

\bibitem[\protect\astroncite{{Min} {\em et~al.\/}}{2007}]{min07}
{Min}, M., {Waters}, L.~B.~F.~M., {de Koter}, A., {Hovenier}, J.~W., {Keller},
  L.~P., \& {Markwick-Kemper}, F. 2007, {\em \aap\/}, {\bf 462}, 667

\bibitem[\protect\astroncite{{Mittal} {\em et~al.\/}}{2015}]{mittal15}
{Mittal}, T., {Chen}, C.~H., {Jang-Condell}, H., {Manoj}, P., {Sargent}, B.~A.,
  {Watson}, D.~M., \& {Lisse}, C.~M. 2015, {\em \apj\/}, {\bf 798}, 87

\bibitem[\protect\astroncite{{Mo{\'o}r} {\em et~al.\/}}{2009}]{moor09}
{Mo{\'o}r}, A., {\em et~al.\/} 2009, {\em \apjl\/}, {\bf 700}, L25

\bibitem[\protect\astroncite{{Mo{\'o}r} {\em et~al.\/}}{2021}]{moor21}
--- 2021, {\em \apj\/}, {\bf 910}, 27

\bibitem[\protect\astroncite{{Morales} {\em et~al.\/}}{2009}]{morales09}
{Morales}, F.~Y., {\em et~al.\/} 2009, {\em \apj\/}, {\bf 699}, 1067

\bibitem[\protect\astroncite{{Morlok} {\em et~al.\/}}{2014}]{morlok14}
{Morlok}, A., {Mason}, A.~B., {Anand}, M., {Lisse}, C.~M., {Bullock}, E.~S., \&
  {Grady}, M.~M. 2014, {\em \icarus\/}, {\bf 239}, 1

\bibitem[\protect\astroncite{Nesvold {\em et~al.\/}}{2016}]{Nesvold_2016}
Nesvold, E.~R., Naoz, S., Vican, L., \& Farr, W.~M. 2016, {\em The
  Astrophysical Journal\/}, {\bf 826}, 19

\bibitem[\protect\astroncite{{Nidever} {\em et~al.\/}}{2002}]{nidever02}
{Nidever}, D.~L., {Marcy}, G.~W., {Butler}, R.~P., {Fischer}, D.~A., \& {Vogt},
  S.~S. 2002, {\em \apjs\/}, {\bf 141}, 503

\bibitem[\protect\astroncite{{Okamoto} {\em et~al.\/}}{2003}]{okamoto03}
{Okamoto}, Y.~K., {\em et~al.\/} 2003, in {\em Instrument Design and
  Performance for Optical/Infrared Ground-based Telescopes\/}, edited by
  M.~{Iye} \& A.~F.~M. {Moorwood}, vol. 4841 of {\em Society of Photo-Optical
  Instrumentation Engineers (SPIE) Conference Series\/},  169--180

\bibitem[\protect\astroncite{{Olofsson} {\em et~al.\/}}{2012}]{olofsson12}
{Olofsson}, J., {Juh{\'a}sz}, A., {Henning}, T., {Mutschke}, H., {Tamanai}, A.,
  {Mo{\'o}r}, A., \& {{\'A}brah{\'a}m}, P. 2012, {\em \aap\/}, {\bf 542}, A90

\bibitem[\protect\astroncite{{Pakhomov} \& {Zhao}}{2013}]{pakhomov13}
{Pakhomov}, Y.~V. \& {Zhao}, G. 2013, {\em \aj\/}, {\bf 146}, 97

\bibitem[\protect\astroncite{{Patience} {\em et~al.\/}}{2002}]{patience02}
{Patience}, J., {Ghez}, A.~M., {Reid}, I.~N., \& {Matthews}, K. 2002, {\em
  \aj\/}, {\bf 123}, 1570

\bibitem[\protect\astroncite{{Payne} {\em et~al.\/}}{2009}]{payne09}
{Payne}, M.~J., {Ford}, E.~B., {Wyatt}, M.~C., \& {Booth}, M. 2009, {\em
  \mnras\/}, {\bf 393}, 1219

\bibitem[\protect\astroncite{{Pilbratt} {\em et~al.\/}}{2010}]{pilbratt10}
{Pilbratt}, G.~L., {\em et~al.\/} 2010, {\em \aap\/}, {\bf 518}, L1

\bibitem[\protect\astroncite{{Poglitsch} {\em et~al.\/}}{2010}]{poglitsch10}
{Poglitsch}, A., {\em et~al.\/} 2010, {\em \aap\/}, {\bf 518}, L2

\bibitem[\protect\astroncite{Price-Whelan {\em
  et~al.\/}}{2017}]{Price-Whelan_2017}
Price-Whelan, A.~M., Hogg, D.~W., Foreman-Mackey, D., \& Rix, H.-W. 2017, {\em
  The Astrophysical Journal\/}, {\bf 837}, 20

\bibitem[\protect\astroncite{{Pu} \& {Wu}}{2015}]{pu15}
{Pu}, B. \& {Wu}, Y. 2015, {\em \apj\/}, {\bf 807}, 44

\bibitem[\protect\astroncite{{Rhee} {\em et~al.\/}}{2007}]{rhee07}
{Rhee}, J.~H., {Song}, I., \& {Zuckerman}, B. 2007, {\em \apj\/}, {\bf 671},
  616

\bibitem[\protect\astroncite{{Rhee} {\em et~al.\/}}{2008}]{rhee08}
--- 2008, {\em \apj\/}, {\bf 675}, 777

\bibitem[\protect\astroncite{{Skrutskie} {\em et~al.\/}}{2006}]{skrutskie06}
{Skrutskie}, M.~F., {\em et~al.\/} 2006, {\em \aj\/}, {\bf 131}, 1163

\bibitem[\protect\astroncite{{Soderblom} {\em et~al.\/}}{2014}]{soderblom14}
{Soderblom}, D.~R., {Hillenbrand}, L.~A., {Jeffries}, R.~D., {Mamajek}, E.~E.,
  \& {Naylor}, T. 2014, in {\em Protostars and Planets VI\/}, edited by
  H.~{Beuther}, R.~S. {Klessen}, C.~P. {Dullemond}, \& T.~{Henning},  219

\bibitem[\protect\astroncite{{Song} {\em et~al.\/}}{2005}]{song05}
{Song}, I., {Zuckerman}, B., {Weinberger}, A.~J., \& {Becklin}, E.~E. 2005,
  {\em \nat\/}, {\bf 436}, 363

\bibitem[\protect\astroncite{{Stassun} {\em et~al.\/}}{2019}]{stassun19}
{Stassun}, K.~G., {\em et~al.\/} 2019, {\em \aj\/}, {\bf 158}, 138

\bibitem[\protect\astroncite{Stauffer {\em et~al.\/}}{1985}]{Stauffer_1985}
Stauffer, J., Hartmann, L., Burnham, J., \& Jones, B. 1985, {\em The
  Astrophysical Journal\/}, {\bf 289}, 247

\bibitem[\protect\astroncite{{Strassmeier} {\em et~al.\/}}{1990}]{strassmeir90}
{Strassmeier}, K.~G., {Fekel}, F.~C., {Bopp}, B.~W., {Dempsey}, R.~C., \&
  {Henry}, G.~W. 1990, {\em \apjs\/}, {\bf 72}, 191

\bibitem[\protect\astroncite{{Su} {\em et~al.\/}}{2020}]{su20}
{Su}, K. Y.~L., {Rieke}, G.~H., {Melis}, C., {Jackson}, A.~P., {Smith}, P.~S.,
  {Meng}, H. Y.~A., \& {G{\'a}sp{\'a}r}, A. 2020, {\em \apj\/}, {\bf 898}, 21

\bibitem[\protect\astroncite{{Su} {\em et~al.\/}}{2009}]{su09}
{Su}, K.~Y.~L., {\em et~al.\/} 2009, {\em \apj\/}, {\bf 705}, 314

\bibitem[\protect\astroncite{{Tajiri} {\em et~al.\/}}{2020}]{tajiri20}
{Tajiri}, T., {\em et~al.\/} 2020, {\em \apjs\/}, {\bf 251}, 18

\bibitem[\protect\astroncite{Vican {\em et~al.\/}}{2016}]{Vican_2016}
Vican, L., Schneider, A., Bryden, G., Melis, C., Zuckerman, B., Rhee, J., \&
  Song, I. 2016, {\em The Astrophysical Journal\/}, {\bf 833}, 263

\bibitem[\protect\astroncite{{Vogt}}{1987}]{vogt87}
{Vogt}, S.~S. 1987, {\em \pasp\/}, {\bf 99}, 1214

\bibitem[\protect\astroncite{{Vogt} {\em et~al.\/}}{2014}]{vogt14}
{Vogt}, S.~S., {\em et~al.\/} 2014, {\em \pasp\/}, {\bf 126}, 359

\bibitem[\protect\astroncite{{Wizinowich}}{2013}]{wizinowich13}
{Wizinowich}, P. 2013, {\em \pasp\/}, {\bf 125}, 798

\bibitem[\protect\astroncite{{Wizinowich} {\em et~al.\/}}{2000}]{wizinowich00}
{Wizinowich}, P., {\em et~al.\/} 2000, {\em \pasp\/}, {\bf 112}, 315

\bibitem[\protect\astroncite{{Wright} {\em et~al.\/}}{2011}]{wright11}
{Wright}, N.~J., {Drake}, J.~J., {Mamajek}, E.~E., \& {Henry}, G.~W. 2011, {\em
  \apj\/}, {\bf 743}, 48

\bibitem[\protect\astroncite{{Yelda} {\em et~al.\/}}{2010}]{yelda10}
{Yelda}, S., {Lu}, J.~R., {Ghez}, A.~M., {Clarkson}, W., {Anderson}, J., {Do},
  T., \& {Matthews}, K. 2010, {\em \apj\/}, {\bf 725}, 331

\bibitem[\protect\astroncite{{Zuckerman}}{2015}]{zuckerman15}
{Zuckerman}, B. 2015, {\em \apj\/}, {\bf 798}, 86

\bibitem[\protect\astroncite{{Zuckerman} {\em et~al.\/}}{2008}]{zuckerman08}
{Zuckerman}, B., {Fekel}, F.~C., {Williamson}, M.~H., {Henry}, G.~W., \&
  {Muno}, M.~P. 2008, {\em \apj\/}, {\bf 688}, 1345

\bibitem[\protect\astroncite{Zuckerman {\em et~al.\/}}{2012}]{Zuckerman_2012}
Zuckerman, B., Melis, C., Rhee, J.~H., Schneider, A., \& Song, I. 2012, {\em
  The Astrophysical Journal\/}, {\bf 752}, 58

\end{thebibliography}
\end{document}